\documentclass[10pt]{article}
\usepackage[T1]{fontenc}
\usepackage[latin9]{inputenc}
\usepackage{amsmath}
\usepackage{amssymb}
\usepackage{graphicx}

\makeatletter
\usepackage{enumitem}		

\usepackage{M11}

\AtBeginDocument{
  
}

\makeatother

\begin{document}
\settitlesize{14.8pt}

\setpreprint{QGASLAB-13-12-01, DMUS-MP-13/21}

\newcommand{\llmbda}{\alpha}
\newcommand{\bpole}{\beta}
\newcommand{\myD}{\Sigma}

\let\oldhash\#
\renewcommand{\#}{\mathbin{\sharp}}

\newcommand{\two}{\mathbin{\natural}}

\let\oldtimes\times
\renewcommand{\times}{\!\oldtimes\!}

\title{Meromorphic Functions and the Topology of Giant Gravitons}

\author{Michael C. Abbott,\alabel{\sharp} Jeff Murugan,\alabel{\sharp}
Andrea Prinsloo\hspace{0.1em}\alabel{\natural} and Nitin Rughoonauth\rlap{\hspace{0.1em}\alabel{\sharp,\flat}}\address[\sharp]{Laboratory
for Quantum Gravity \& Strings, Department of Mathematics \& Applied
Mathematics,\\
 University of Cape Town, Rondebosch 7701, South Africa }\address[\natural]{Department
of Mathematics, University of Surrey, Guildford, GU2 7XH, England}\address[\flat]{Max-Planck-Institut
f\"ur Physik, F\"ohringer Ring 6, 80805 Munich, Germany, and \\
Arnold-Sommerfeld-Center für Theoretische Physik, \\
LMU München, Theresienstraße 37, 80333 Munich, Germany}\address{michael.abbott@uct.ac.za,
jeff@nassp.uct.ac.za, a.prinsloo@surrey.ac.uk, nitincr@gmail.com}}

\date{17 December 2013}

\maketitle
\begin{abstract}
Using Mikhailov's map from holomorphic functions to supersymmetric
D3-brane solutions, we show how to construct giant gravitons in $AdS_{5}\times S^{5}$
with toroidal topologies. In the $\tfrac{1}{4}$-BPS sector we show
that these are always of the form $\#^{K}(S^{2}\times S^{1})$, and
in the limit in which this becomes a set of $m+n$ perpendicular spherical
giants re-connected near to their intersections, we find $K$ in terms
of $m,n$. In the $\tfrac{1}{8}$-BPS sector we find a similar class
of solutions. 
\end{abstract}

\section{Introduction}

The best understood sector of AdS/CFT \cite{Maldacena:1997re} concerns
closed fundamental strings, which in the gauge theory are single-trace
operators of length much less than $N$ \cite{Beisert:2010jr}. D-branes
are heavier objects corresponding to operators of length order $N$,
the most studied of which are spherical branes known as giant gravitons
\cite{McGreevy:2000cw}, dual to determinant operators and generalisations
known as Schur polynomials \cite{Balasubramanian:2001nh,Corley:2001zk,Koch:2009gq}.

The worldsheet of any closed string state is a cylinder $\mathbb{R}\times S^{1}$,
but for D3-branes any $\mathbb{R}\times\mathcal{M}$ is in principle
possible, with $\mathcal{M}$ a closed 3-manifold. Such manifolds
can have considerably more complicated topology than closed 1-manifolds,
and it would be fascinating to understand the emergence of topology
from the dual SYM operators. Motivated by this, the goal of this letter
is to explore what topologies occur in giants with a given amount
of supersymmetry. We are particularly interested in solutions created
by a localised modification of a set of intersecting spherical giant
gravitons, as this seems the most tractable limit. 


We begin by recalling the map given by Mikhailov \cite{Mikhailov:2000ya,Mikhailov:2002wx}:
Any analytic function $f:\mathbb{C}^{3}\to\mathbb{C}$ defines a supersymmetric
D3-brane solution in $\mathbb{R}\times S^{5}\subset AdS_{5}\times S^{5}$
as the surface 
\begin{equation}
f(e^{-it}Z_{1},\: e^{-it}Z_{2},\: e^{-it}Z_{3})=0,\qquad\smash{\sum_{i}\left|Z_{i}\right|^{2}=1}\label{eq:moving-surface-from-f}
\end{equation}
where $Z_{i}=r_{i}\, e^{i\phi_{i}}$ are the 3 complex embedding co-ordinates
for $S^{5}$. The degree of supersymmetry is given by the number of
arguments: $f(Z_{1})$ gives a $\tfrac{1}{2}$-BPS solution, $f(Z_{1},Z_{2})$
$\tfrac{1}{4}$-BPS, and $f(Z_{1},Z_{2},Z_{3})$ gives an $\tfrac{1}{8}$-BPS
solution. 

The usual sphere giant graviton is given in this language by 
\begin{equation}
f(Z_{1})=Z_{1}-\llmbda.\label{eq:f-simple}
\end{equation}
This constrains $Z_{1}$ completely, so the worldvolume (at a given
time) is the $S^{3}$ parameterised by $Z_{2}$ and $Z_{3}$ subject
to $\left|Z_{2}\right|^{2}+\left|Z_{3}\right|^{2}=1-\llmbda^{2}$.
The only time-evolution is that it rotates in the $Z_{1}$ plane;
the maximal giant graviton has $\llmbda=0$ and is thus stationary,
while in the opposite limit $\llmbda\to1$ the brane collapses down
to a point particle on a lightlike trajectory. A function $f(Z_{1})$
with several zeros will lead to a number of concentric spherical giants. 


We refer to \eqref{eq:f-simple} as the case $(1,0,0)$: one $Z_{1}$
giant. The next section studies the effect of adding to this terms
depending on $Z_{2}$, and then takes a limit in which these give
$n$ intersecting $Z_{2}$ giants: cases $(1,n,0)$. After that we
consider arbitrarily many intersecting $Z_{1}$ and $Z_{2}$ giants,
cases $(m,n,0)$ (section \ref{sec:Class-mn0}), and finally the addition
also of $Z_{3}$ giants (section \ref{sec:Eighth-BPS}). We give a
concise statement of our results in section \ref{sec:Conclusion}.

\section{Quarter-BPS Class $(1,n,0)$\label{sec:Quater-BPS-Class-1n0}}

To begin constructing topologically nontrivial solutions using Mikhailov's
method, in this section we add to the spherical giant's $f(Z_{1})$
a meromorphic function of $Z_{2}$. Consider first the function 
\begin{equation}
f(Z_{1},Z_{2})=Z_{1}-\llmbda+\frac{\epsilon}{Z_{2}}.\label{eq:f-one-pole}
\end{equation}
We may assume $\llmbda,\epsilon>0$ and, since the motion of the brane
is rigid, we need only discuss its topology at time $t=0$. 

Let us parameterise the D3-brane worldvolume by the $\phi_{3}$ circle
and some portion of the $Z_{2}$ plane. (Solving $f=0$ fixes $Z_{1}$
in terms of $Z_{2}$, and $\sum_{i}r_{i}^{2}=1$ fixes $r_{3}$.)
We can read off the topology of the brane from the topology of the
area of the $Z_{2}$ plane thus covered: let $\myD$ be the area where
$r_{3}\geq0$. For example, the spherical giant graviton \eqref{eq:f-simple}
clearly has for $\myD$ the disk $\left|Z_{2}\right|\leq1-\llmbda^{2}$. 

\begin{figure}
\centering  \includegraphics[width=30mm]{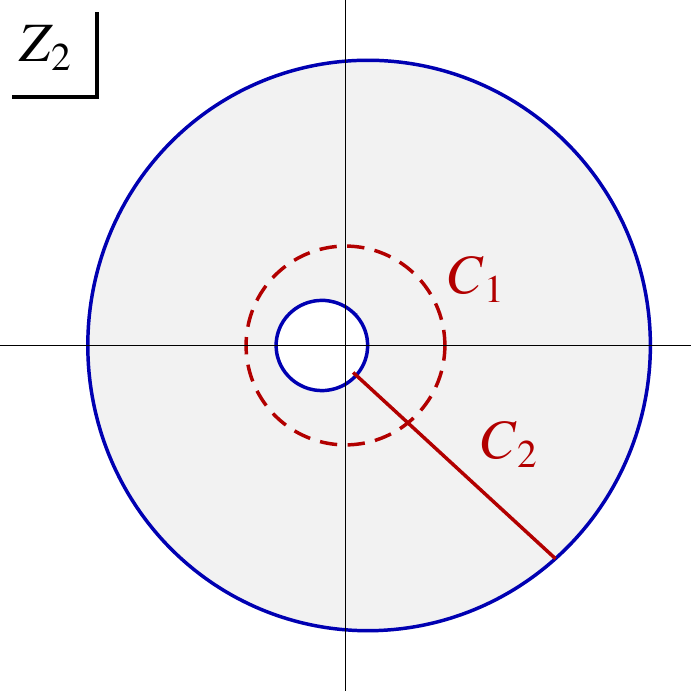} \hspace{5mm}\includegraphics[width=30mm]{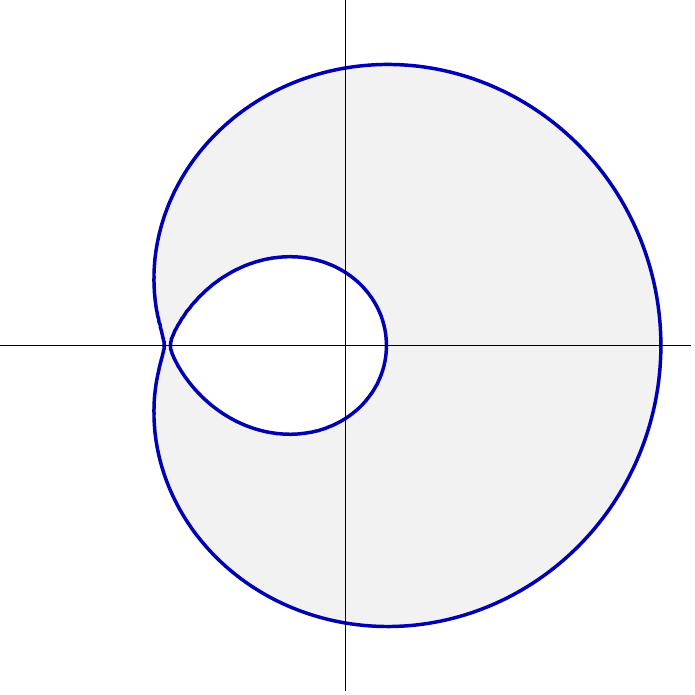}
\hspace{5mm}\includegraphics[width=30mm]{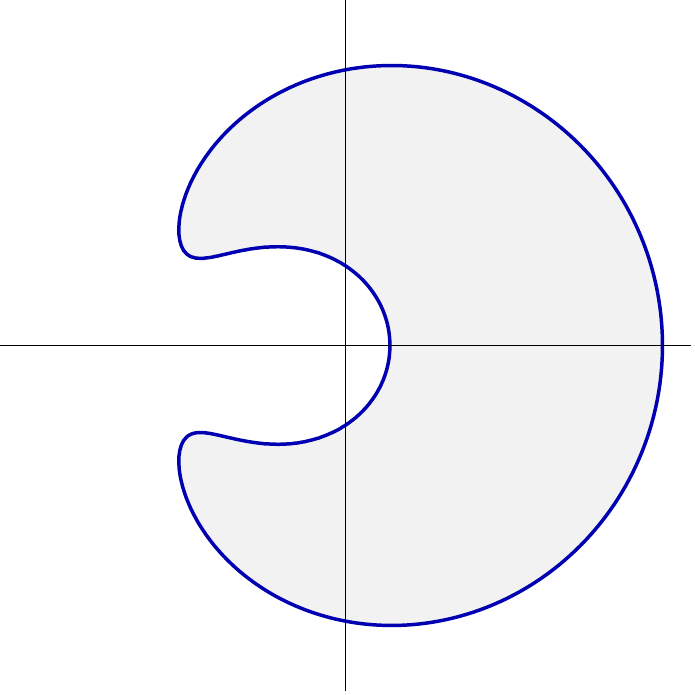} 

\caption[Fake caption]{Plots showing the area $\myD$ of the $Z_{2}$ plane which is covered
by the D3-brane specified by \eqref{eq:f-one-pole}. Increasing the
residue, we progress from a torus $\mathcal{M}=S^{2}\times S^{1}$
via the critical case to a deformed $S^{3}$. Parameters are $\llmbda=\tfrac{1}{2}$
and $\epsilon=\tfrac{1}{10},\:0.1844,\:\tfrac{1}{5}$.\label{fig:Plots-dD-one-pole}
}
\end{figure}

The effect of turning on the pole is to make a hole in the base space
$\myD$, thus increasing its genus (see figure \ref{fig:Plots-dD-one-pole}).
This may be understood by saying that in a neighbourhood of the pole,
the term $\epsilon/Z_{2}$ is so large that there are no solutions
$\left|Z_{1}\right|\leq1$. Notice immediately that this means that
the pole itself is not on the worldvolume of the brane. 

To analyse this more carefully, it is easy to show using \eqref{eq:f-one-pole}
that $\myD$ is given by 
\begin{equation}
r_{2}^{\,4}+r_{2}^{\,2}(\llmbda^{2}-1)+\epsilon^{2}\leq2\epsilon\llmbda\, r_{2}\cos\phi_{2}.\label{eq:my-r-theta-eq}
\end{equation}
Drawing graphs of the left- and right-hand sides in terms of $r_{2}^{\,2}$,
when $\epsilon=0$ certainly there are two intersections. Increasing
$\epsilon$, there is a range $0<\epsilon<\epsilon_{\text{crit}}$
in which there are two intersections $r_{2}>0$ for all $\phi_{2}$,
followed by a range $\epsilon_{\text{crit}}<\epsilon<\epsilon_{\text{max}}$
in which there are two intersections at $\phi_{2}=0$ but not at $\phi_{2}=\pi$.
For larger $\epsilon$ there are no intersections. This progression
is shown in figure \ref{fig:Plots-dD-one-pole}. 

For $\epsilon<\epsilon_{\text{crit}}$ the topology of the brane is
$S^{2}\times S^{1}$. The incontractible cycle $C_{1}$ (more or less
the $\phi_{2}$ circle) is the $S^{1}$ factor, while a radial line
in $\myD$ gives the $S^{2}$ factor --- this is an interval fibered
with the $\phi_{3}$ circle, which shrinks to zero at either end.
It is marked $C_{2}$ in the figure.

\begin{figure}
\centering \includegraphics[width=30mm]{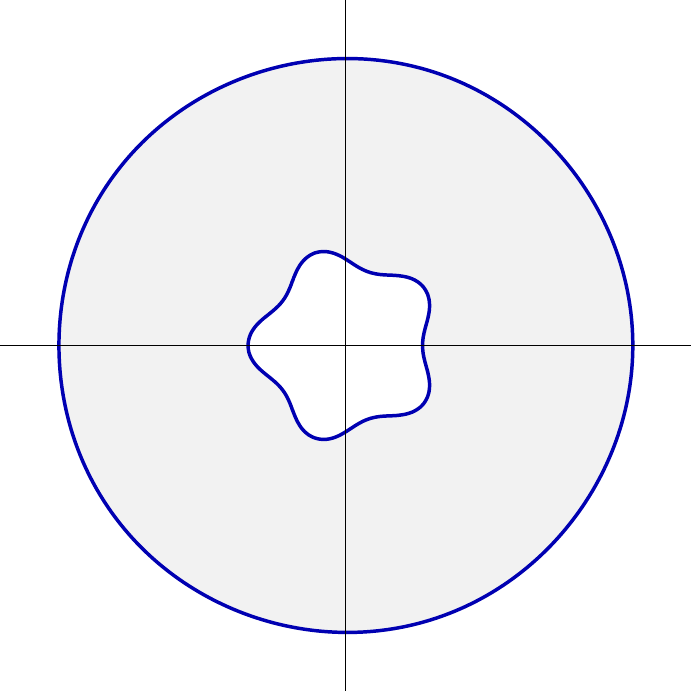} \hspace{5mm}\includegraphics[width=30mm]{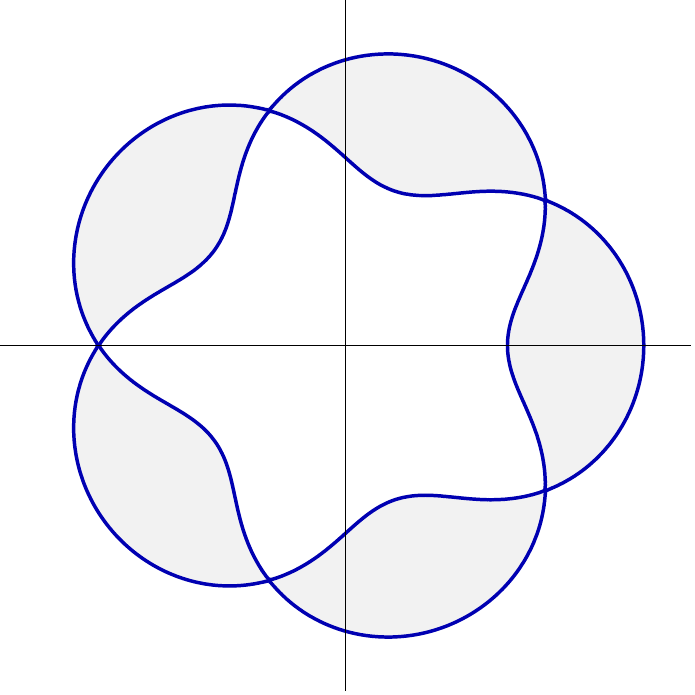}
\hspace{5mm}\includegraphics[width=30mm]{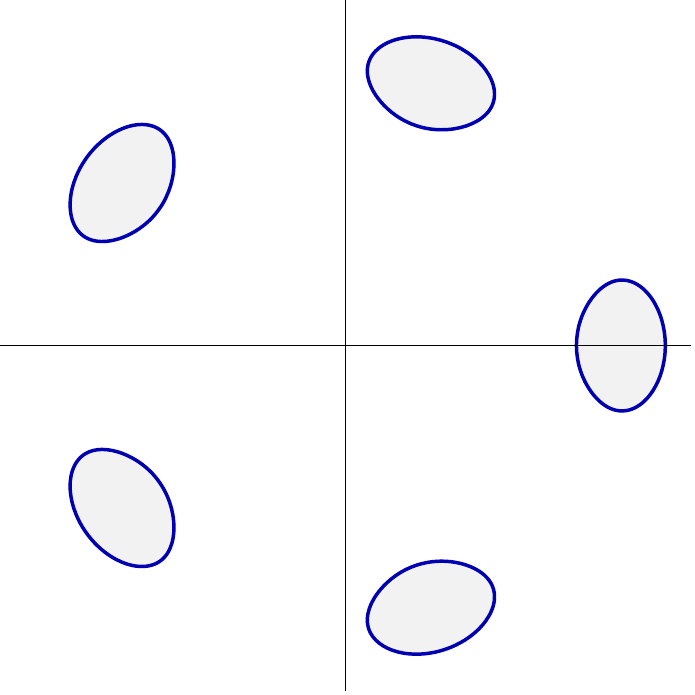} 

\caption[Fake caption]{Plots showing $\myD$ with one quintuple pole, \eqref{eq:f-n-fold-pole}
with $N=5$. Increasing $\epsilon$ we pass from the torus on the
left to five spheres on the right; the middle picture is close to
$\epsilon_{\mathrm{crit}}$. Parameters are $\llmbda=\tfrac{1}{2}$
and $\epsilon=\frac{1}{1000},0.03833,\frac{1}{5}$. \label{fig:Plots-dD-higher-pole}
}
\end{figure}

The simplest generalisation is to consider a higher-order pole: 
\begin{equation}
f(Z_{1},Z_{2})=Z_{1}-\llmbda+\frac{\epsilon}{(Z_{2})^{N}}.\label{eq:f-n-fold-pole}
\end{equation}
This leads to the same topology as the single pole, for small $\epsilon$,
but the geometry has a symmetry $Z_{2}\to e^{i2\pi/N}Z_{2}$. Because
of this, in the regime $\epsilon_{\mathrm{crit}}<\epsilon<\epsilon_{\mathrm{max}}$
there will be $N$ separate (deformed) 3-spheres. Figure \ref{fig:Plots-dD-higher-pole}
shows the case of $N=5$. 

\begin{figure}
\centering  \includegraphics[width=32mm]{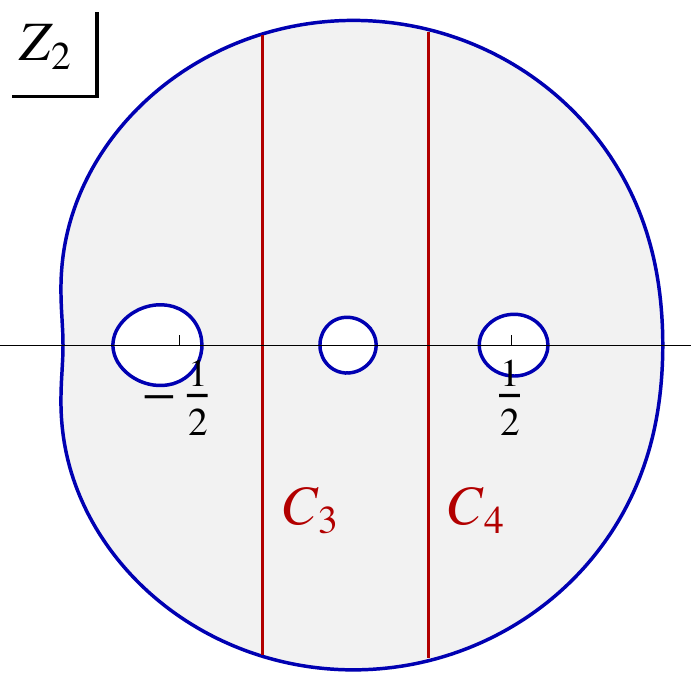} \hspace{3mm}\includegraphics[width=32mm]{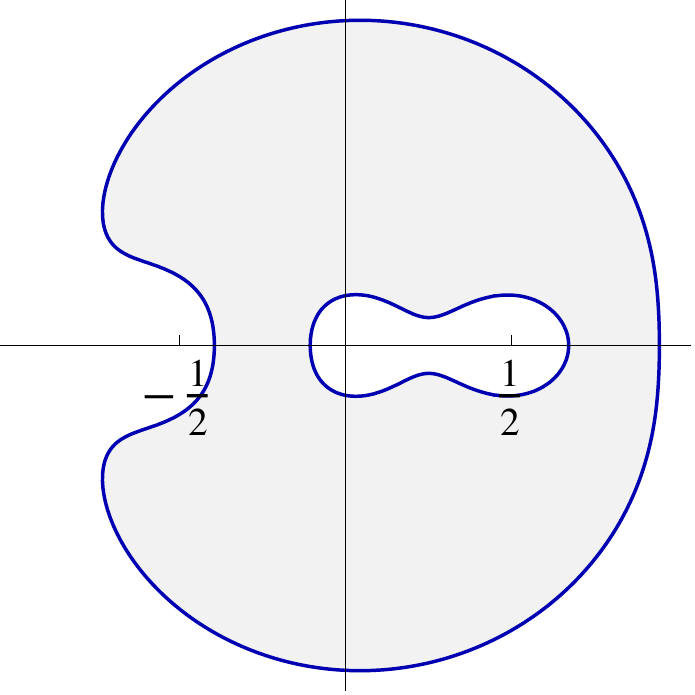}
\hspace{3mm}\includegraphics[width=32mm]{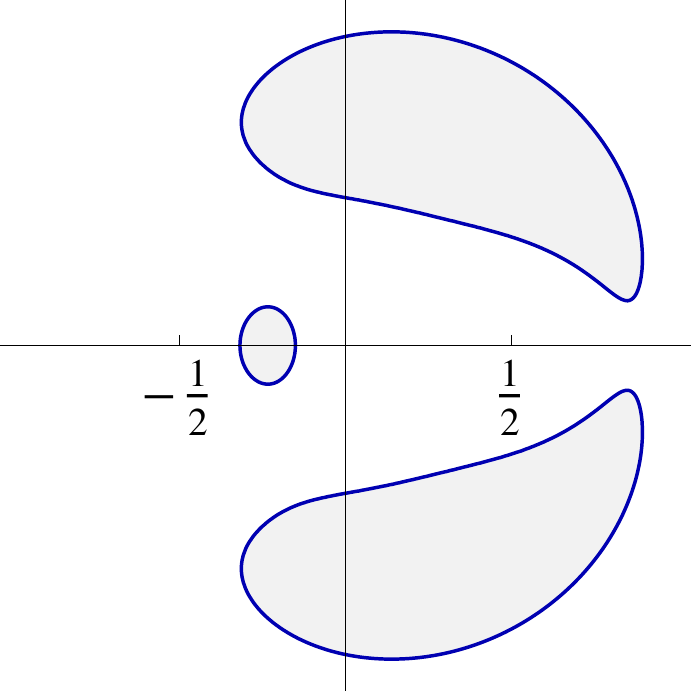} 

\caption[Fake caption]{Plots showing $\myD$ for the case $(1,3,0)$, using \eqref{eq:f-many-poles}
with three poles at $\bpole_{1}=-\tfrac{1}{2}$, $\bpole_{2}=0$ and
$\bpole_{3}=\tfrac{1}{2}$. The residues are $\epsilon,\epsilon,-\epsilon$
respectively, with $\epsilon$ increasing from $\tfrac{1}{12}$ (left,
giving $\mathcal{M}=\#^{3}(S^{2}\times S^{1})$) to $\tfrac{1}{7}$
(centre, giving $S^{2}\times S^{1}$) to $\tfrac{1}{3}$ (right, $\sqcup_{3}S^{3}$),
and $\llmbda=\tfrac{1}{2}$. Notice that the holes in $\myD$ formed
by residues of opposite signs attract, while those of the same sign
repel. The lines $C_{3},C_{4}$ in $\myD$ each lift to a separating
$S^{2}$ in $\mathcal{M}$. \label{fig:Plots-dD-three-poles} }
\end{figure}

We can also consider several poles: 
\begin{equation}
f(Z_{1},Z_{2})=Z_{1}-\llmbda+\sum_{j=1}^{n}\frac{\epsilon_{j}}{Z_{2}-\bpole_{j}}.\label{eq:f-many-poles}
\end{equation}
For small enough $\epsilon_{j}$ the analysis very close to each pole
will be similar to that for one pole: expanding in $r_{2\beta}=\left|Z_{2}-\bpole_{j}\right|$
will give us \eqref{eq:my-r-theta-eq} plus terms higher order in
$r_{2\beta}$. Thus for any set of $n$ poles (located at $\bpole_{j}$
such that $\llmbda^{2}+\left|\bpole_{j}\right|^{2}<1$) there exist
residues $\epsilon_{j}\neq0$ such that $\myD$ is a disk with $n$
holes. Cutting $\myD$ along lines such as $C_{3},C_{4}$ in figure
\ref{fig:Plots-dD-three-poles}, so that each hole is isolated, we
see that the resulting topology is a connected sum%
\footnote{Recall that the notion of a connected sum is this: If cutting a 3-manifold
$M$ along an $S^{2}$ separates the manifold into $M_{1}'\sqcup M_{2}'$,
and $M_{i}$ is $M_{i}'$ with a 3-ball glued to its boundary, then
we write $M=M_{1}\#M_{2}$. The sphere is the identity in the sense
$M=M\#S^{3}$. Every (oriented closed connected) 3-manifold has a
unique decomposition as a sum of prime manifolds, primeness meaning
that every separating $S^{2}$ bounds a ball. 

The connected sum of 2-manifolds, which we write $\two$, is defined
by similarly cutting along $S^{1}$. This gives rise to the genus
classification of surfaces $S^{2},T^{2},T^{2}\two T^{2},\two^{g}T^{2}$. %
} 
\begin{equation}
\mathcal{M}=\#^{n}(S^{2}\times S^{1}).\label{eq:case-1n0-M}
\end{equation}
Note that all poles are outside $\myD$, so everywhere on the worldvolume
the function $f$ is analytic. 

With multiple poles the progression as we increase $\epsilon_{j}$
can be quite complicated, and can produce several disconnected pieces.
The case of poles at $Z_{2}=0,\pm\tfrac{1}{2}$ is shown in figure
\ref{fig:Plots-dD-three-poles}. Notice that the holes created by
$\epsilon_{2}/Z_{2}$ and $\epsilon_{3}/(Z_{2}-\beta_{3})$ merge
with each other in the middle picture. The same effect can be produced
by moving them together at fixed $\epsilon$: when $\beta_{3}\to0$
these two approach \eqref{eq:f-n-fold-pole}. We study this kind of
degeneration limit extensively below. 

Returning for a moment to our simplest case \eqref{eq:f-one-pole},
there are two more distinct degeneration limits given by the two cycles
shown: 
\begin{itemize}
\item As $\epsilon\to\epsilon_{\text{crit }}$, $C_{2}$ becomes small in
a throat which is locally $S^{2}\times\mathbb{R}$. This is the right
geometry to be interpreted as the effect of some strings pinching
the brane into a torus. In the case $\llmbda=0$ the brane becomes
an equally thin tube all around, approaching the circular spinning
string solution 
\begin{equation}
Z_{1}=\tfrac{1}{\sqrt{2}}e^{i(t+\sigma)},\qquad Z_{2}=\tfrac{1}{\sqrt{2}}e^{i(t-\sigma)},\qquad Z_{3}=0.\label{eq:spinning-string}
\end{equation}
The toroidal brane may thus be thought of as an embiggened circular
string. But note that all D3-branes described by \eqref{eq:moving-surface-from-f}
carry no worldsheet electric field $F_{01}$, and thus the string
solution here is not an F-string. 
\item As $\epsilon\to0$, instead $C_{1}$ becomes small in a throat locally
$S^{1}\times\mathbb{R}^{2}$. Here it is useful to think of the $\epsilon\neq0$
function \eqref{eq:f-one-pole} not as the addition of a meromorphic
term to \eqref{eq:f-simple} but (multiplying through by the denominator)
as a the addition of a small term to a factorised polynomial. That
is, $f=Z_{1}Z_{2}+\epsilon$ describes exactly the same D3-brane as
\eqref{eq:f-one-pole}, but in the limit $\epsilon\to0$ more obviously
approaches $f=Z_{1}Z_{2}$, which is a pair of intersecting maximal
sphere giants. \medskip \\ The effect of infinitesimal $\epsilon$
is localised near to their interesection: At $Z_{2}\neq0$ it changes
$Z_{1}=0$ to $Z_{1}=\epsilon/Z_{2}$, perturbing the $Z_{1}$ giant
smoothly away from maximality (and likewise $Z_{2}$). But the effect
near to $Z_{1}=Z_{2}=0$ is not smooth, as $\sqcup_{2}S^{3}$ is re-connected
so as to give topologically $S^{2}\times S^{1}$. 
\end{itemize}
For three poles in \eqref{eq:f-many-poles} (and taking the numbers
used in figure \ref{fig:Plots-dD-three-poles}) the view suggested
by the limit $\epsilon\to0$ is of four intersecting branes, and hence
we refer to this as the case $(1,3,0)$: 
\begin{align*}
f(Z_{1},Z_{2}) & =(Z_{1}-\tfrac{1}{2})(Z_{2}+\tfrac{1}{2})Z_{2}(Z_{2}-\tfrac{1}{2})+\epsilon\:(Z_{2}^{2}-Z_{2}-1)\\
 & \approx(Z_{1}-\tfrac{1}{2})Z_{2}(Z_{2}^{2}-\tfrac{1}{4})+\epsilon'
\end{align*}
The three $Z_{2}$ branes intersect the $Z_{1}$ brane at different
places, and since $\epsilon\neq0$ modifies the solution appreciably
only near to the intersection, it is natural that the effect of several
$Z_{2}$ branes is very simply related to the effect of one, \eqref{eq:M-case-1n1}.
In the next section we study more general cases in this limit, allowing
also several $Z_{1}$ branes.

\section{Class $(m,n,0)$\label{sec:Class-mn0}}

\newcommand{\torus}{{\raisebox{-1.2mm}{\includegraphics[width=7mm]{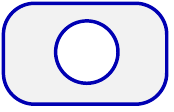}}}}
\newcommand{\smalltorus}{\smash{\raisebox{-0.5mm}{\includegraphics[width=5.5mm]{torus.pdf}}}}

\newcommand{\tube}{\hspace{-5mm}\smash{\raisebox{0.1mm}{\includegraphics[width=7mm]{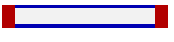}}}\hspace{-5mm}}
\newcommand{\vtube}{\hspace{0.5mm}{\raisebox{-1mm}{\includegraphics[height=4.5mm]{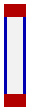}}}\hspace{0.5mm}}
\newcommand{\twotube}{\rlap{\raisebox{1mm}{\tube}}\raisebox{-1mm}{\tube}}

\newcommand{\tubespace}{\smash{\raisebox{0mm}{\includegraphics[width=7mm]{tube.pdf}}}}

Figure \ref{fig:Plots-dD-three-poles} above shows $\myD$ for the
case (1,3,0). Let us now analyse the case (3,1,0), which must be equivalent.
The simplest example is
\begin{equation}
f=(Z_{1}^{3}-\alpha^{3})Z_{2}+\epsilon.\label{eq:f-one-triple}
\end{equation}
Solving for $Z_{1}=\sqrt[3]{\alpha^{3}-\epsilon/Z_{2}}$, if we again
call the area of the $Z_{2}$ plane occupied by the solution $\myD$,
this is now a Riemann surface with three sheets. Each sheet is a disk
with one hole, and the branch cut (from $Z_{2}=0$ to $\epsilon/\alpha^{3}$)
runs from a point in the hole to a point inside $\myD$. Drawing the
connections as in figure \ref{fig:About-the-3,1,0}, it is clear that
$\myD$ has three holes, thus we recover $\mathcal{M}=\#^{3}(S^{2}\times S^{1})$
as desired. The same procedure works equally well for branch cuts
of any order. 

\begin{figure}
\newcommand{\cutsone}{\raisebox{-15mm}{\includegraphics[scale=0.4]{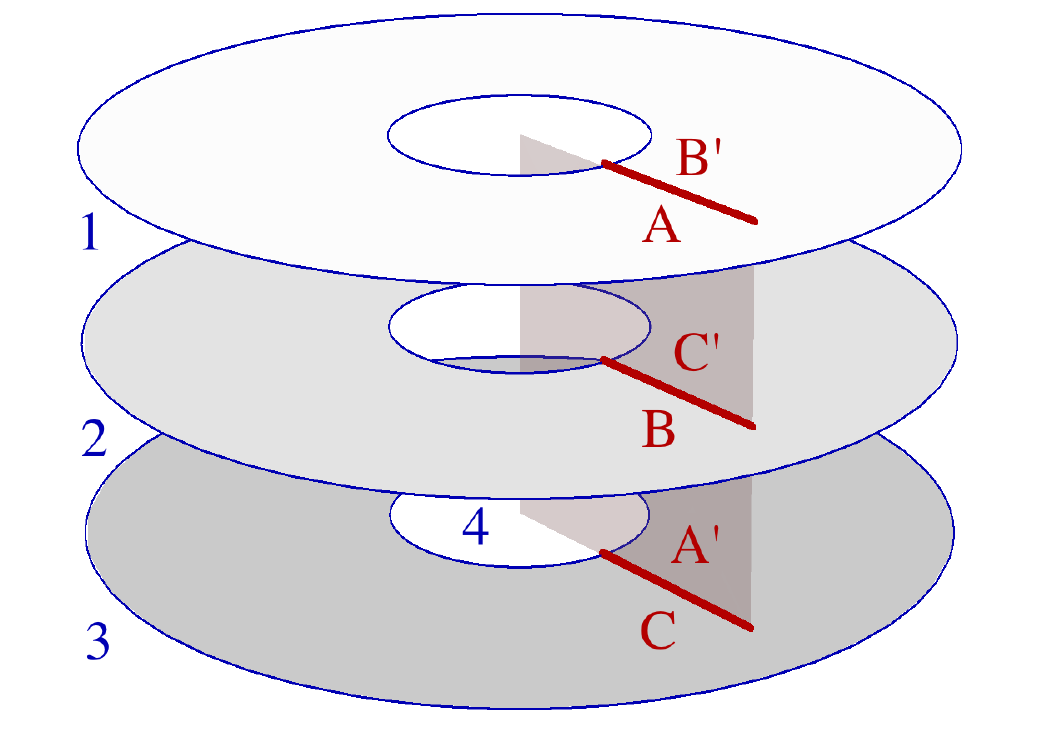}}}
\newcommand{\cutstwo}{\raisebox{-15mm}{\includegraphics[scale=0.4]{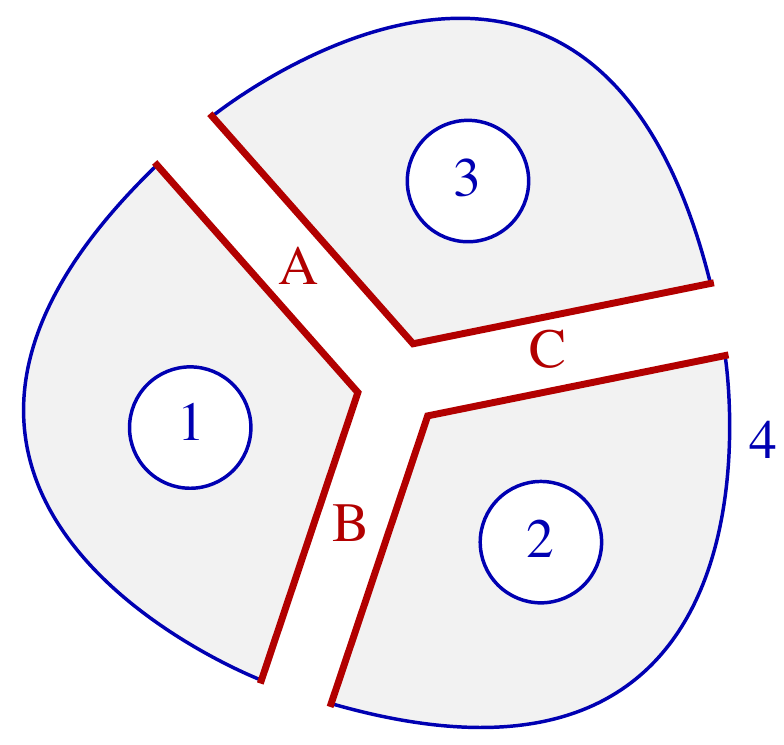}}} 
\[
\cutsone=\quad\cutstwo\quad=\quad\begin{array}{c}
\torus\\
\vtube\\
\torus\\
\vtube\\
\torus
\end{array}
\]

\caption[Fake caption]{Branch cut for the $(3,1,0)$ case \eqref{eq:f-one-triple}. Each
sheet of $\myD$ on the left can be turned inside-out to give a wedge
as shown. (The numbers label boundary components.) Glueing these back
together, the result is a disk with three holes, drawn schematically
at the right, and equivalent to figure \ref{fig:Plots-dD-three-poles}a.
(Note that the cuts labelled $A=A'$ etc. are \emph{not} the $S^{2}$
glue lines of the connected sum.) \label{fig:About-the-3,1,0} }
\end{figure}

Before moving on to a new case, we make the following observation:
The connected sum of two 3-manifolds $\mathcal{M},\mathcal{N}$ can
be regarded as the effect of connecting a tube $S^{2}\times I$ (i.e.
$S^{3}$ with two punctures) between any point on one and any point
on the other: \newcommand{\tinypic}[1]{\smash{\raisebox{-3mm}{\includegraphics[scale=0.16]{#1}}}}
\[
\mathcal{M}\#\mathcal{N}=\mathcal{M}\#S^{3}\#\mathcal{N}.\qquad\qquad\tinypic{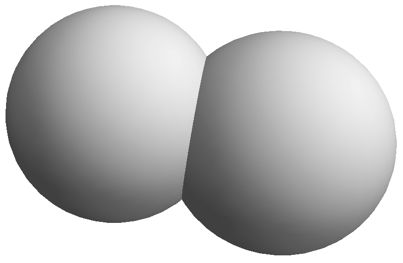}=\tinypic{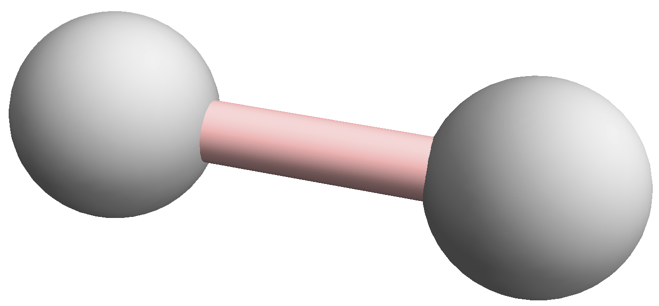}\hspace{-1cm}
\]
Connecting a similar tube between two points on the \emph{same} manifold
$\mathcal{M}$ instead has the effect of adding one term $S^{2}\times S^{1}$:
\begin{equation}
\mathcal{M}+(S^{2}\times I\text{ handle})=\mathcal{M}\#(S^{2}\times S^{1}).\qquad\qquad\tinypic{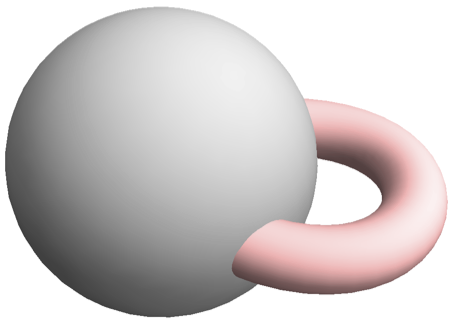}=\tinypic{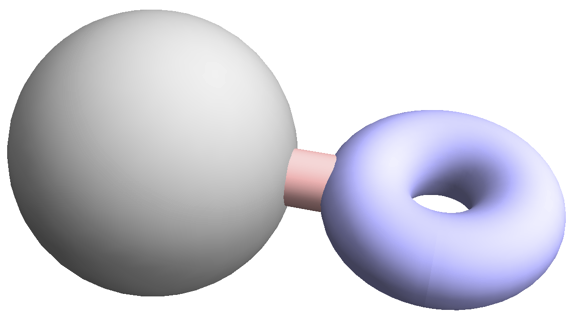}\hspace{-1cm}\label{eq:handle-addition}
\end{equation}
In the notation of the figures, where $\smalltorus=S^{2}\times S^{1}$
and $\tubespace=S^{2}\times I\text{ tube}$, this reads\newcommand{\myspace}{\hspace{5mm}}
\[
\torus\myspace\twotube\myspace\torus\;=\;\torus\myspace\tube\myspace\torus\myspace\tube\myspace\torus\;=\;\#^{3}\torus.
\]

\begin{figure}
\newcommand{\temp}[1]{\raisebox{-17mm}{\includegraphics[width=32mm]{#1.pdf}}}
\[
\temp{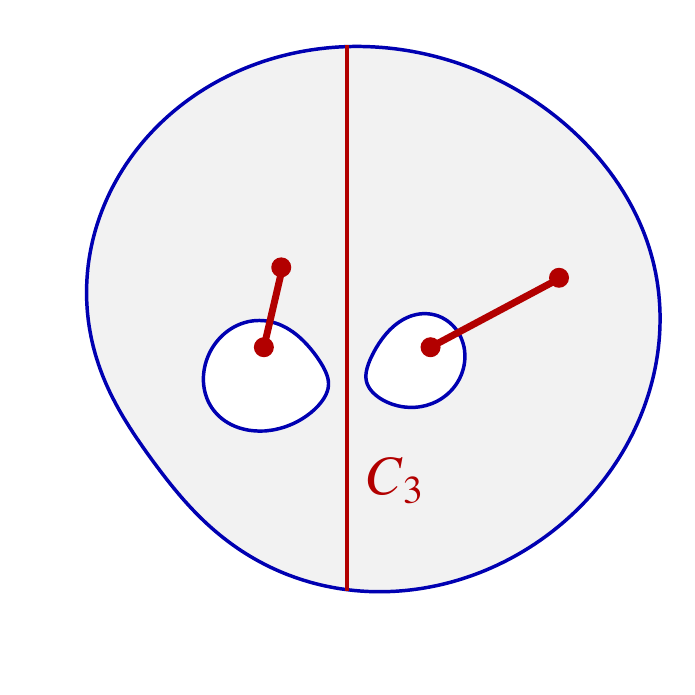}=\begin{array}{ccc}
\torus & \tube & \torus\\
\vtube &  & \vtube\\
\torus & \tube & \torus
\end{array}\quad\text{ or }\quad\temp{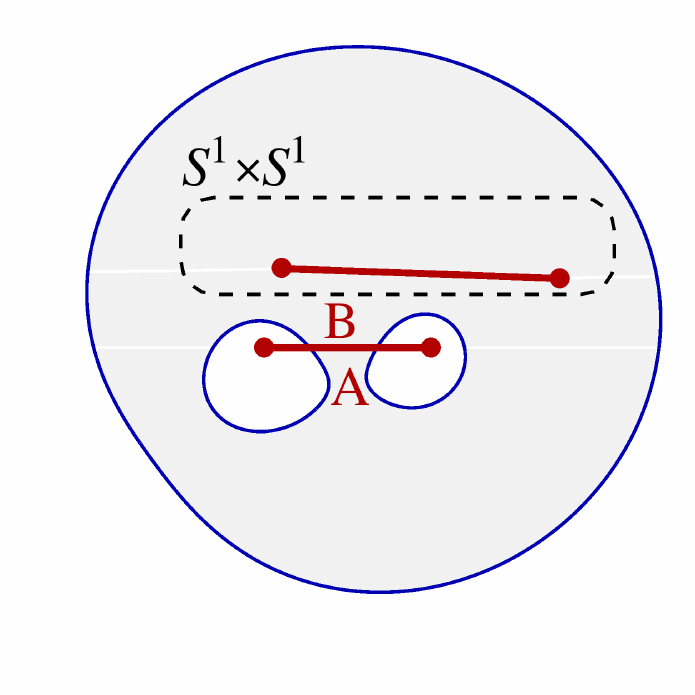}=\temp{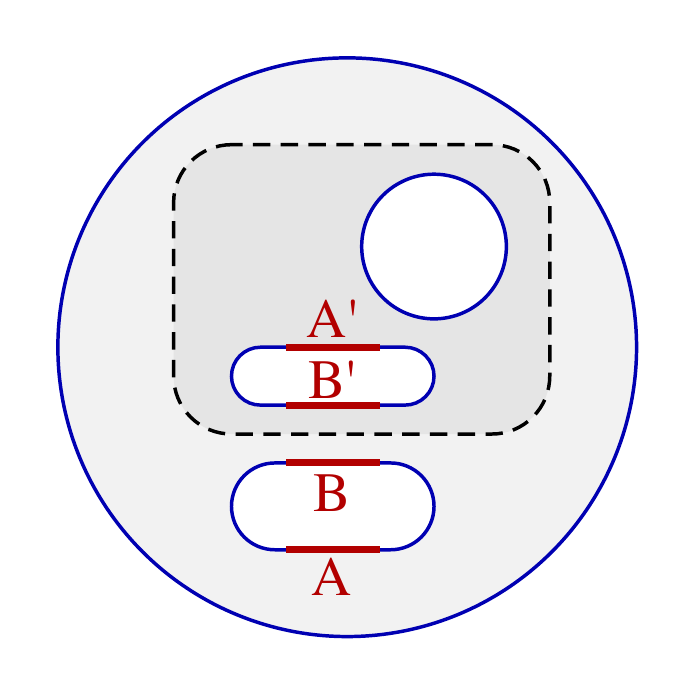}\vspace{-4mm}
\]

\caption[Fake caption]{Branch cuts for case $(2,2,0)$. The first plot shows two square-root
cuts each of which can be treated as in figure \ref{fig:About-the-3,1,0},
giving the vertical connections; the horizontal connections are from
the glue line $C_{3}$. The second plot shows the same branch points
connected the other way. In this case we can pull the lower sheet
of $\Sigma$ through the cut to obtain the figure on the right. (This
happens within the dashed line. The circular boundary component was
the outer boundary of the lower sheet.) Now $\Sigma$ is a disk with
three holes plus two handles, giving the same topology. \label{fig:Branch-cuts-for-220}}
\end{figure}

Now consider the case $(2,2,0)$, starting with
\begin{equation}
f=(Z_{1}^{2}-\alpha^{2})(Z_{2}^{2}-\beta^{2})+\epsilon.\label{eq:f-220-simple}
\end{equation}
Clearly $\myD$ has two sheets, each with two holes, connected by
a pair of branch cuts. To analyse this, we can split each sheet along
surfaces $C_{i}$ like those used in the previous section --- see
figure \ref{fig:Branch-cuts-for-220}. This gives two copies of the
$(2,1,0)$ case, $\#^{2}\smalltorus$, connected in two places. Re-connecting
along $C_{3}$ on the upper sheet gives the connected sum of the two
pieces, and re-connecting the lower sheet adds a handle of the type
just discussed. In all we get
\[
\mathcal{M}=\left[\#^{2}(S^{2}\times S^{1})\vphantom{1^{1^{1}}}\right]\#\left[\#^{2}(S^{2}\times S^{1})\vphantom{1^{1^{1}}}\right]+(S^{2}\times I\text{ handle})=\#^{5}(S^{2}\times S^{1}).
\]
Figure \ref{fig:Branch-cuts-for-220} shows this procedure. Instead
of \eqref{eq:f-220-simple} it uses $f=(Z_{1}^{2}-\alpha^{2})+\epsilon/(Z_{2}-\beta)+i\epsilon/(Z_{2}+\beta)$,
with $\alpha=\tfrac{1}{2}$, $\beta=\tfrac{1}{4}$ and $\epsilon=\tfrac{1}{9}$,
to have a convenient arrangement of branch points. It also shows an
alternative argument to check that the choice of how we draw the branch
cuts does not matter. 

Generalising to the case $(m,n,0)$, the topology is 
\begin{equation}
\mathcal{M}=\#^{K}(S^{2}\times S^{1}),\qquad K=mn+(m-1)(n-1).\label{eq:M-1/4-BPS-mn0-single}
\end{equation}
The counting comes from drawing a grid of $\smalltorus$ and connecting
horizontally (as in figure \ref{fig:Plots-dD-three-poles}) and vertically
(as figure \ref{fig:About-the-3,1,0}). 

\begin{figure}
\newcommand{\smalltemp}[1]{\raisebox{-12mm}{\includegraphics[width=28mm]{#1.pdf}}}
\[
\smalltemp{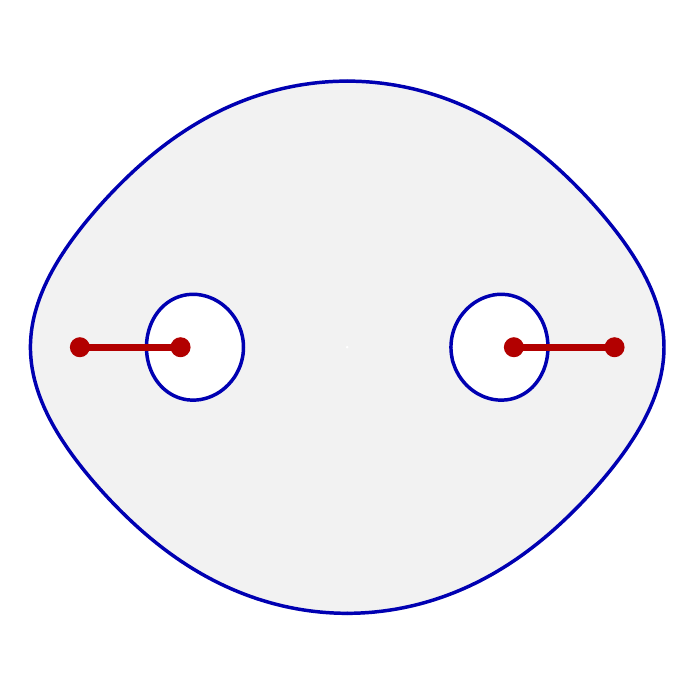}\quad\longrightarrow\quad\underset{\text{small }\alpha}{\smalltemp{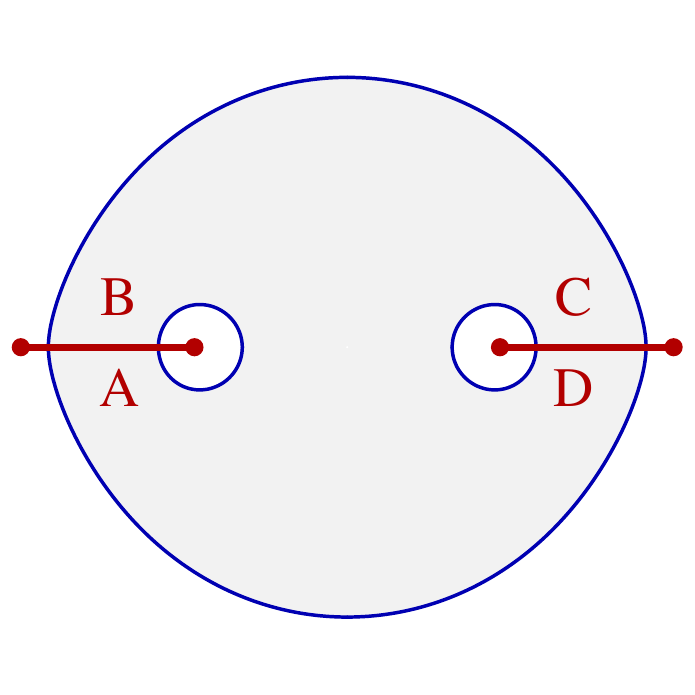}}\quad\text{ or }\quad\underset{\text{small }\beta}{\smalltemp{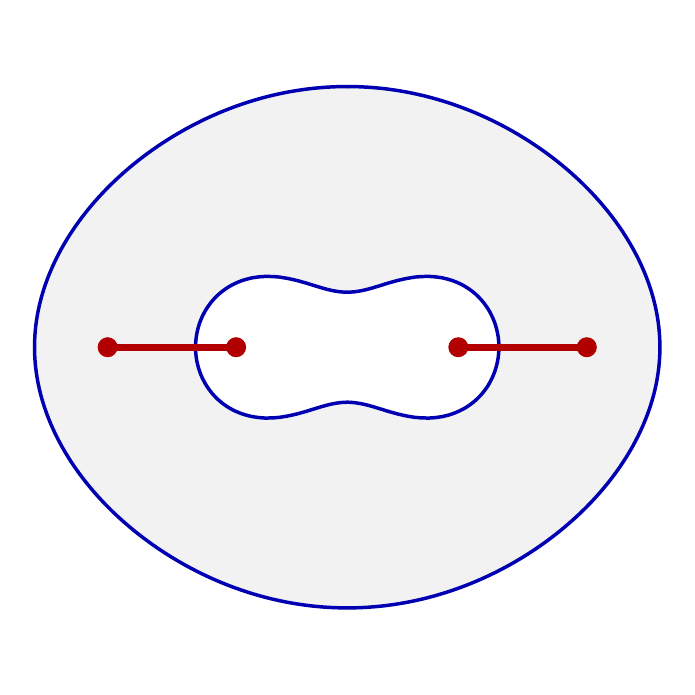}}=\begin{array}{c}
\torus\\
\vtube\vtube\\
\torus
\end{array}\vspace{-3mm}
\]

\caption[Fake caption]{Degeneration limits of the $(2,2,0)$ case \eqref{eq:f-220-simple},
drawing always the upper sheet of $\myD$. For the central picture
(small $\beta$) we have two disks connected by 4 handles (labelled
$A\ldots D$), while for the right-hand picture (small $\alpha$)
we have 2 tori connected by 2 handles. (The initial picture is $\alpha=\beta=\tfrac{1}{2}$,
$\delta=\tfrac{1}{10}$, and for each limit drawn ``small'' means
$\tfrac{1}{3}$.)\label{fig:Degeneration-limits-of-220}\\ \smash{\raisebox{50mm}{\hspace{115mm}\includegraphics[width=40mm]{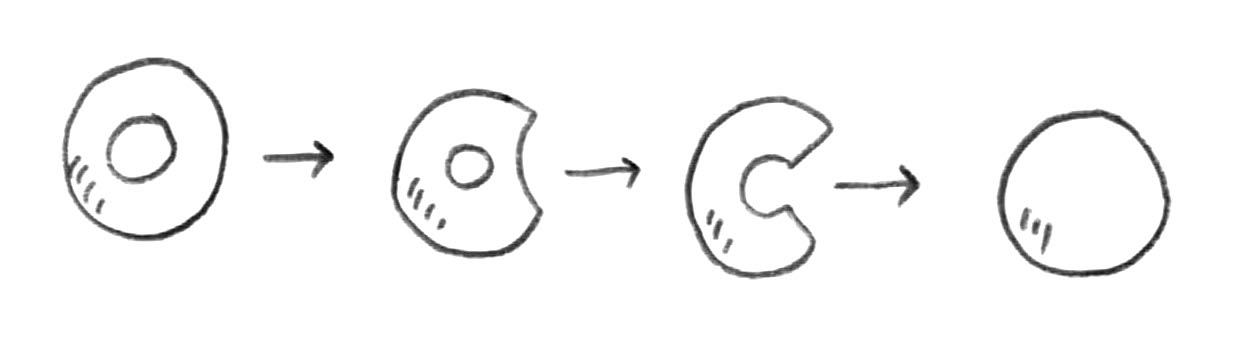}}}\vspace{-4mm}}
\end{figure}

So far we have assumed that the $m+n$ intersecting branes are all
at distinct positions, or in other words we considered only single
poles. In the last section, allowing instead higher-order poles \eqref{eq:f-n-fold-pole}
did not change the topology, but this is no longer true here. We can
investigate this by moving poles to co-incide. There are two ways
to do this in \eqref{eq:f-220-simple}, taking either $\alpha\to0$
or $\beta\to0$, and these must be equivalent. Solving for $Z_{1}$,
the branch points are located at 
\[
Z_{2}=\pm\beta,\quad\pm\sqrt{\beta^{2}+\epsilon/\alpha^{2}}.
\]
For small $\epsilon$ the second pair are inside $\myD$, giving the
analysis above. But (holding $\epsilon$ fixed and) taking the limit
$\alpha\to0$, these move off to infinity, giving cuts all the way
across $\myD$. In the limit $\beta\to0$, instead the holes in $\myD$
merge into one (as happened in figure \ref{fig:Plots-dD-three-poles}).
Both situations are drawn in figure \ref{fig:Degeneration-limits-of-220},
and each leads to $\mathcal{M}=\#^{3}(S^{2}\times S^{1})$. 

It is now clear how to treat any set of poles of any order. Write
$n$ for the number of separated poles in $Z_{2}$, and $N$ for their
total order. (Two double poles thus give $N=4$, $n=2$, as does one
single and one triple pole.) Similarly write $M\geq m$ for poles
in $Z_{1}$. Drawing an $m\times n$ grid of $\smalltorus$, the number
of vertical connections is $M$, and horizontal $N$ --- see figure
\ref{fig:About-degeneration...}. Then counting the holes we get 
\[
\mathcal{M}=\#^{K}(S^{2}\times S^{1}),\qquad K=1+M(n-1)+N(m-1).\tag{\ref{eq:final-1/4-BPS-M}}
\]
This change from \eqref{eq:M-1/4-BPS-mn0-single} is a result of holes
in $\myD$ merging with each other. We learned in section \ref{sec:Quater-BPS-Class-1n0}
that the effect of increasing $\epsilon$ is similar. Thus we expect
that, for a completely general $\tfrac{1}{4}$-BPS giant, the topology
will still be $\#^{K}(S^{2}\times S^{1})$ for some $K$. 

\begin{figure}
\[
\underset{\vphantom{1^{1^{\frac{1}{1}}}}M=m=N=n=3,\; K=13}{\begin{array}{ccccc}
\torus & \tube & \torus & \tube & \torus\\
\vtube &  & \vtube &  & \vtube\\
\torus & \tube & \torus & \tube & \torus\\
\vtube &  & \vtube &  & \vtube\\
\torus & \tube & \torus & \tube & \torus
\end{array}}\quad\underset{\beta_{1}\to\beta_{2}}{\longrightarrow}\quad\underset{\vphantom{1^{1^{\frac{1}{1}}}}N=3,\; n=2,\; K=10}{\begin{array}{ccc}
\torus & \tube & \torus\\
\vtube\vtube &  & \vtube\\
\torus & \tube & \torus\\
\vtube\vtube &  & \vtube\\
\torus & \tube & \torus
\end{array}}\quad\underset{\alpha_{2}\to\alpha_{3}}{\longrightarrow}\quad\underset{\vphantom{1^{1^{\dfrac{1^{1}}{1}}}}M=N=3,\; m=n=2,\; K=7}{\begin{array}{ccc}
\torus & \tube & \torus\\
\vtube\vtube &  & \vtube\\
\torus & \twotube & \torus
\end{array}}
\]

\caption[Fake caption]{Degeneration of the case $(3,3,0)$. Starting with three distinct
poles at $Z_{1}=\alpha_{i}$ and three at $Z_{2}=\beta_{j}$, we first
allow two $\beta$ poles to merge into a double pole, and then two
$\alpha$ poles likewise. $\mathcal{M}$ is given by \eqref{eq:final-1/4-BPS-M}
with the numbers shown.  \label{fig:About-degeneration...}  }
\end{figure}

\section{Eighth-BPS\label{sec:Eighth-BPS}}

We now turn wish to turn on at least one intersecting $Z_{3}$ giant,
and take a similar small-$\epsilon$ limit. Using what we have learned,
we can immediately treat all cases $(1,n,1)$ together. Consider 
\begin{equation}
f=Z_{1}Z_{3}+\epsilon\: h(Z_{2})\label{eq:f-1n1}
\end{equation}
where $h$ is a function with $n$ poles. Clearly $f=0$ fixes $\phi_{+}=\phi_{1}+\phi_{3}$
and the product $r_{1}r_{3}$ in terms of $Z_{2}$, while $\phi_{-}=\phi_{1}-\phi_{3}$
is unconstrained. We can solve for $r_{1}$ and $r_{3}$ as separate
functions of $Z_{2}$ by writing $\sum_{i}r_{i}^{2}=1$ as 
\[
(r_{1}\pm r_{3})^{2}=1-r_{2}^{2}\pm2\, r_{1}r_{3}\equiv H_{\pm}
\]
and substituting in $r_{1}r_{3}=\epsilon\left|h\right|$. This gives
\[
(r_{1},r_{3})=\tfrac{1}{2}\left(\sqrt{H_{+}}\pm\sqrt{H_{-}},\:\sqrt{H_{+}}\mp\sqrt{H_{-}}\right).
\]
At a point $Z_{2}$ for which $H_{-}\geq0$ there are two solutions
(coalescing when $H_{-}=0$), while for $H_{-}<0$ there are none.
Notice that where $h$ has a pole, $H_{-}\to-\infty$, and thus the
neighbourhood of such points will be excluded. (And for small $\epsilon$,
each such hole will be small.) Define $\myD$ to be two copies of
the area of the $Z_{2}$ plane for which $H_{-}\geq0$, sewn up along
the boundary. For one pole this is a simple torus, while for $n$
poles  $\Sigma=\two^{n}T^{2}$. 

In the $\tfrac{1}{4}$-BPS case we always had an $S^{1}$ fibred over
$\Sigma$, shrinking to a point on $\partial\Sigma$. Fitting with
the fact that this boundary is now empty, the $\phi_{-}$ circle here
never shrinks to a point. To check, note that the metric is 
\[
ds^{2}=\sum_{i=1}^{3}\left(dr_{i}^{2}+r_{i}^{2}d\phi_{i}^{2}\right)=2(r_{1}^{2}+r_{3}^{2})d\phi_{-}^{2}+\ldots.
\]
Thus for the length of the $\phi_{-}$ circle to be zero we need $r_{1}=r_{3}=0$,
which implies $r_{2}=1$, and this is never part of $\Sigma$. We
conclude that the topology is 
\begin{equation}
\mathcal{M}=(\two^{n}T^{2})\times S^{1}.\label{eq:M-case-1n1}
\end{equation}
Note that all such three-manifolds are prime --- the connected sum
here is the two-dimensional one, and to separate the manifold (nontrivially)
we must cut along $T^{2}$ surfaces. 

\begin{figure}
\centering \vspace{-4mm}\raisebox{3mm}{\includegraphics[width=43mm]{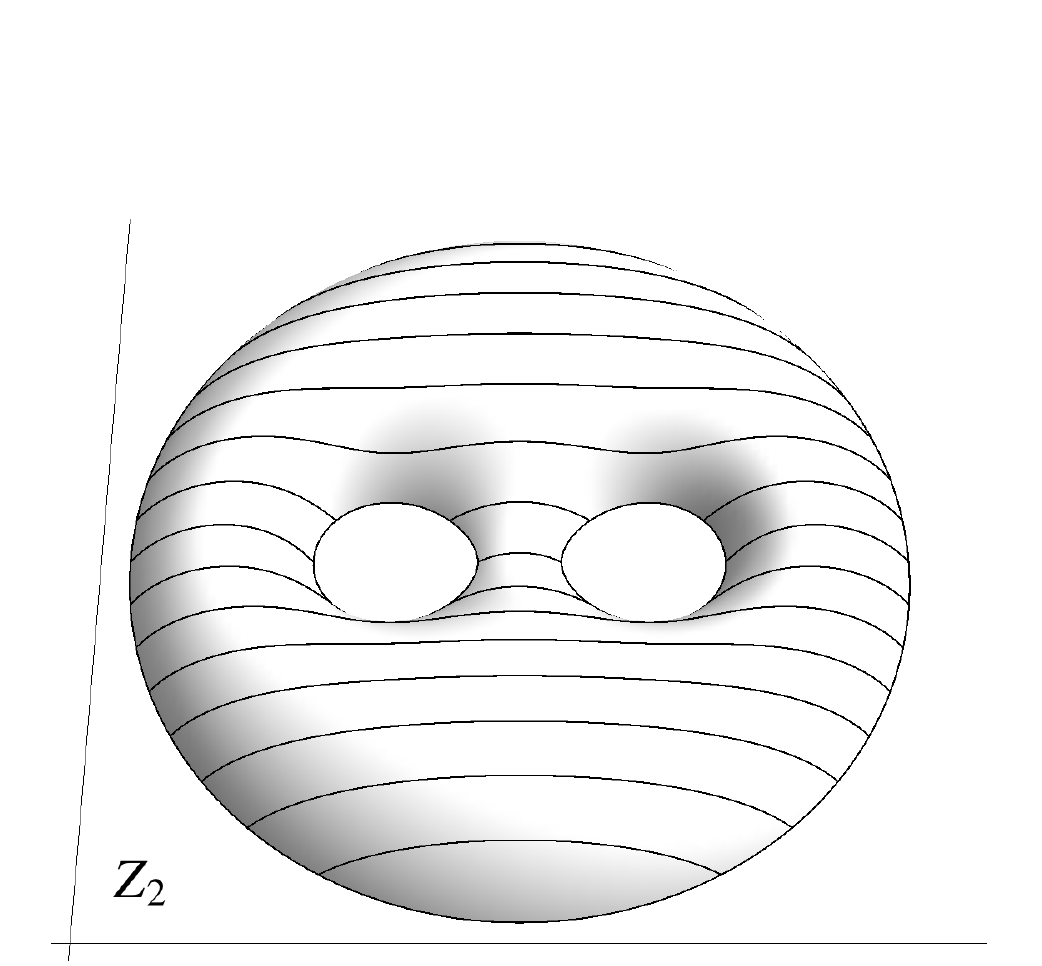}}\hspace{5mm}\includegraphics[width=40mm]{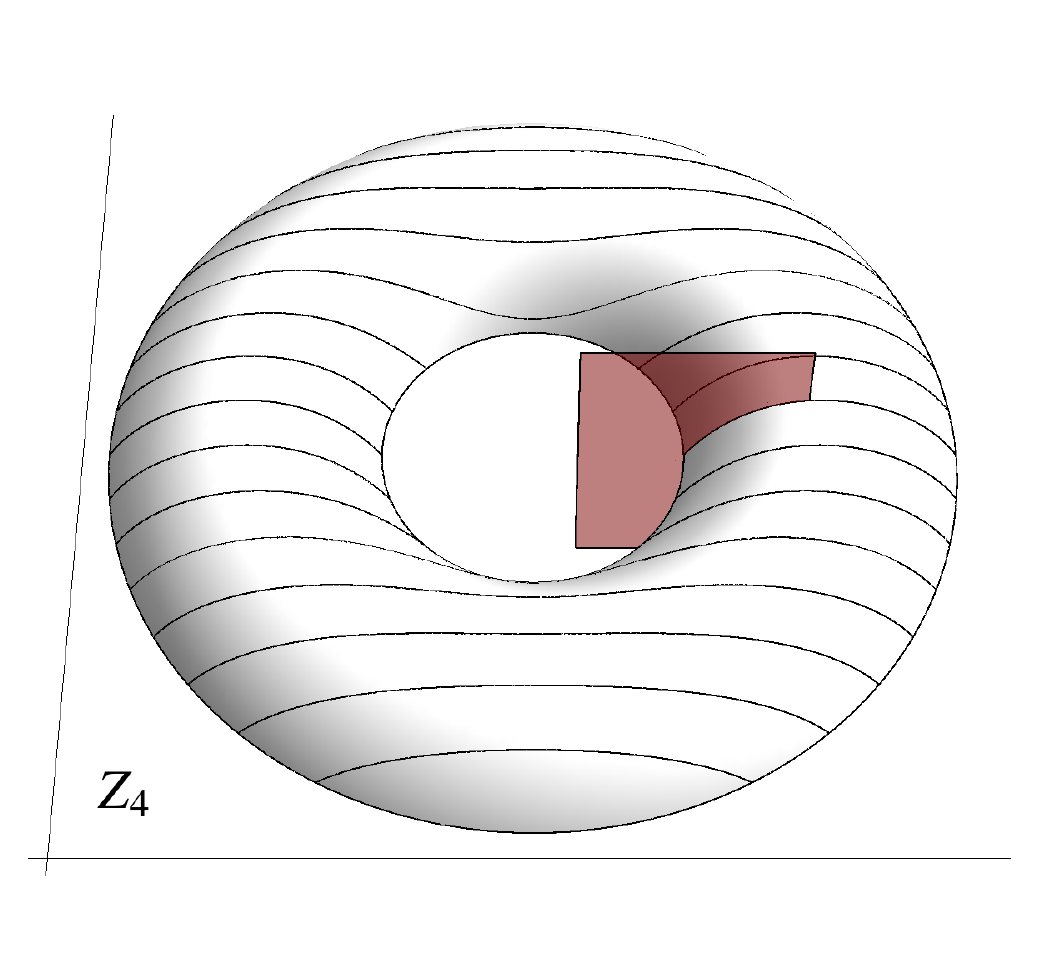}

\caption[Fake caption]{Plots for the $\tfrac{1}{8}$-BPS case $(1,2,1)$. The first picture
shows $\Sigma$ dawn as two points fibered over $Z_{2}$. The second
shows the upper of two sheets in the $Z_{4}=Z_{1}Z_{3}$ plane, which
are connected by a branch cut drawn in red. \label{fig:Plots-for-1/8-BPS}}
\end{figure}

While the area of the $Z_{2}$ plane involved here is not identical
to that for the $\tfrac{1}{4}$-BPS case of section \ref{sec:Quater-BPS-Class-1n0},
the spirit is clearly very similar: each pole increases the genus
of the base space $\myD$. It is natural to ask how much of our analysis
of section \ref{sec:Class-mn0} still holds. To understand this we
begin by re-analysing $f=Z_{1}(Z_{2}^{2}-\beta^{2})Z_{3}+\epsilon$,
the case $(1,2,1)$. Solving for $Z_{2}$ we get 
\[
Z_{2}=\pm\frac{\beta\sqrt{Z_{1}Z_{3}-\epsilon/\beta^{2}}}{\sqrt{Z_{1}Z_{3}}}.
\]
It is natural to think of this as having a branch cut in the $Z_{4}=Z_{1}Z_{3}$
plane. Fixing our position on this plane fixes $Z_{2}$ up to a choice
of sheets, after which we still have (at a generic point) two solutions
$(r_{1},r_{3})$. Figure \ref{fig:Plots-for-1/8-BPS} shows the upper-sheet
part of $\myD$ which, when glued along the branch cut drawn, gives
a double torus, and thus the same topology as before. The angle in
the $S^{1}$ factor is still $\phi_{1}-\phi_{3}$.

One extension beyond \eqref{eq:f-1n1} is now fairly obvious. If we
consider%
\footnote{Note aside that we could likewise consider $\tfrac{1}{4}$-BPS solutions
of the form $f=(Z_{1}-\alpha)(Z_{4}-\gamma)+\epsilon'$. For small
$\gamma$ this can give a double torus $\#^{2}(S^{2}\times S^{1})$,
but small $\epsilon'$ here does not guarantee that $\epsilon$ in
\eqref{eq:final-f-product} is small. %
} 
\[
f=1+\sum_{k=1}^{m}\frac{\epsilon'_{k}}{Z_{1}Z_{3}-\gamma_{k}}+\sum_{j=1}^{n}\frac{\epsilon_{j}}{Z_{2}-\beta_{j}}
\]
then we can re-use all the $\tfrac{1}{4}$-BPS analysis. Just as we
replaced $\myD$ of figure \ref{fig:Plots-dD-three-poles} with two
copies glued along their edges to get \eqref{eq:M-case-1n1}, similarly
replace $\myD$ of figures \ref{fig:Branch-cuts-for-220}, \ref{fig:About-degeneration...}
with their closed cousins. We get $(\two^{K}T^{2})\times S^{1}$ with
the same $K$ as before. 

But it is essential here that $f$ contains only the product $Z_{1}Z_{3}$.
The more natural class $(m,n,1)$ of solutions
\[
f=Z_{3}+\sum_{i=1}^{m}\frac{\epsilon'_{i}}{Z_{1}-\alpha_{i}}+\sum_{j=1}^{n}\frac{\epsilon_{j}}{Z_{2}-\beta_{j}}
\]
will break the $S^{1}$ symmetry in \eqref{eq:M-case-1n1}. We leave
the analysis of this, and completely general $\tfrac{1}{8}$-BPS cases,
for future work.

\section{Conclusion\label{sec:Conclusion}}

The main result of this letter is as follows:
\begin{quote}
Let $g(Z_{1},Z_{2})$ be a meromorphic function with $m$ distinct
poles at $Z_{1}=\alpha_{i}$, and write $M$ for the number of poles
counting multiplicity. Similarly let $n$ and $N$ count the poles
at $Z_{2}=\beta_{j}$. We require $m,n\geq1$, and $\left|\alpha_{i}\right|^{2}+\left|\beta_{j}\right|^{2}\leq1\;\forall i,j$.%
\footnote{We could weaken this condition to allow for cases where not every
pair of branes intersect in the limit $\epsilon\to0$; this has the
effect of deleting some nodes from the corners of the lattice shown
in figure \ref{fig:About-degeneration...}, and thus reducing $K$,
but not otherwise changing the topology. %
} Consider the $\tfrac{1}{4}$-BPS giant described by 
\begin{equation}
f(Z_{1},Z_{2})=1+\epsilon\: g(Z_{1},Z_{2}).\label{eq:final-f-g}
\end{equation}
For sufficiently small $\epsilon$, this has topology specified by
the prime decomposition 
\begin{equation}
\mathcal{M}=\#^{K}(S^{2}\times S^{1}),\qquad K=1+M(n-1)+N(m-1).\label{eq:final-1/4-BPS-M}
\end{equation}
As $\epsilon$ is increased, generically%
\footnote{This is true if the residues of $g$ are constants, in which case
$\sum_{i}K_{i}\leq K$ and $L+L'\leq MN$ in \eqref{eq:disjoint-lots}.
But if the numerator of $g$ is of sufficiently high order then $K$
may increase. %
} $K$ will decrease, and the brane may break up into several disjoint
pieces. All pieces are either spheres or connected sums of $(S^{2}\times S^{1})$:
\begin{equation}
\mathcal{M}=\bigsqcup_{i}^{L}\#^{K_{i}}(S^{2}\times S^{1})\bigsqcup_{j}^{L'}S^{3}.\label{eq:disjoint-lots}
\end{equation}
 
\end{quote}
We found it convenient to deal with a function $f$ with poles, which
in some sense repel the base space $\myD$ thus creating holes in
the brane.%
\footnote{The poles are thus never on the worldvolume, so $f$ is locally analytic,
which is enough to guarantee a solution to the equations of motion
from \eqref{eq:moving-surface-from-f}. %
} The same solutions can equivalently be specified by polynomial functions
of the form%
\footnote{Here $M=\sum_{i}\mu_{i}$ and $N=\sum_{j}\nu_{j}$.%
} 
\begin{equation}
f(Z_{1},Z_{2})=\prod_{i=1}^{m}(Z_{1}-\alpha_{i})^{\mu_{i}}\prod_{j=1}^{n}(Z_{2}-\beta_{j})^{\nu_{j}}+\epsilon\:\text{poly}(Z_{1},Z_{2}).\label{eq:final-f-product}
\end{equation}
Clearly $\epsilon=0$ gives a factorised $f$ and thus a set of intersecting
spherical giants \eqref{eq:f-simple}. The effect of small $\epsilon$
is to suppress all but the simplest kind of interactions: the topology
is unchanged when the last term here is replaced by a constant. Nevertheless
what we have observed is that the effect of increasing $\epsilon$
is quite simple: the tori degenerate (reducing $K$) and ultimately
split into disjoint spheres. Thus we believe that \eqref{eq:disjoint-lots}
applies to generic polynomial functions $f(Z_{1},Z_{2})$. 

For $\tfrac{1}{8}$-BPS geometries we have more limited results. The
generalisation which can be treated by borrowing much of the analysis
from above is 
\[
f(Z_{1},Z_{2},Z_{3})=1+\epsilon\: g(Z_{4},Z_{2}),\qquad Z_{4}=Z_{1}Z_{3}
\]
with $g$ defined as in \eqref{eq:final-f-g}. For small enough $\epsilon$
the resulting topology is%
\footnote{From the topologies written down here it is trivial to obtain the
homology groups. The Betti numbers are: 
\begin{align*}
\tfrac{1}{4}\text{-BPS}:\qquad b_{0} & =b_{3}=1,\qquad b_{1}=b_{2}=K\\
\tfrac{1}{8}\text{-BPS}:\qquad b_{0} & =b_{3}=1,\qquad b_{1}=b_{2}=2K+1
\end{align*}
For the case $(1,1,0)$, the generators of $H_{1}$ and $H_{2}$ are
cycles $C_{1}$ and $C_{2}$ in figure \ref{fig:Plots-dD-one-pole}.
It is easy to draw similar cycles in figure \ref{fig:Plots-dD-three-poles}'s
case $(1,3,0)$. Similar cycles drawn in figure \ref{fig:Plots-for-1/8-BPS}'s
case $(1,2,1)$ will all be 1-cycles. %
} 
\[
\mathcal{M}=\left[\two^{K}(S^{1}\times S^{1})\right]\times S^{1}
\]
where $K$ is as in \eqref{eq:final-1/4-BPS-M}. Notice that none
of these topologies can occur in the $\tfrac{1}{4}$-BPS case. Generalising
this to allow other combinations of $Z_{3}$ branes (such as \eqref{eq:final-f-product}
with $\prod_{k}(Z_{3}-\gamma_{k})$ inserted) is an open problem.
But it seems clear that the form of $\mathcal{M}$ will change, and
in particular will not have an overall $S^{1}$ factor. 


While our focus in this letter has been entirely on the classical
membranes described by \eqref{eq:moving-surface-from-f}, a detailed
quantisation of the moduli space of Mikhailov solutions was carried
out in \cite{Biswas:2006tj}, and used to draw conclusions about the
spectrum of $\tfrac{1}{8}$-BPS states in $\mathcal{N}=4$ SYM. It
would be of great interest to pursue the relationship between these
results and ours. 

This work forms part of a larger research program aimed at understanding
how local and global properties of spacetime are encoded in gauge
theory. For recent work in this direction, see \cite{Pasukonis:2010rv,Koch:2012ck,Pasukonis:2012zj}
and \cite{Berenstein:2013md} and references therein. In this context,
it would be nice to see how the toplogies studied here emerge in operators
dual to Mikhailov's giants in SYM. There too, perhaps our small-$\epsilon$
limit is likely to be the tractable one. 


\subsection*{Acknowledgements}

We thank David Berenstein, Robert de Mello Koch, Jan Gutowski, Konstadinos
Sfetsos, Jonathan Shock and Alessandro Torielli for discussions. 

M.C.A. is supported by a UCT URC postdoctoral fellowship. J.M. acknowledges
support from the NRF of South Africa under the HCDE and IPRR programs.
N.R. is supported by a DAAD-AIMS scholarship, and a DAAD short-term
research fellowship.


\appendix

\bibliographystyle{my-JHEP-4}
\bibliography{/Users/me/Documents/Papers/complete-library-processed,complete-library-processed}

\end{document}